\documentclass[12pt,preprint]{aastex}

\slugcomment{Accepted for publication in ApJ}

\begin{document}

\title{MECI: A Method for Eclipsing Component Identification}

\author{Jonathan Devor and David Charbonneau\footnote{Alfred P. Sloan Research Fellow}}

\affil{Harvard-Smithsonian Center for Astrophysics, 60 Garden Street, Cambridge, MA 02138}

\email{jdevor@cfa.harvard.edu}

\begin{abstract}
We describe an automated method for assigning the most probable
physical parameters to the components of an eclipsing binary,
using only its photometric light curve and combined colors. With
traditional methods, one attempts to optimize a multi-parameter
model over many iterations, so as to minimize the chi-squared
value. We suggest an alternative method, where one selects pairs
of coeval stars from a set of theoretical stellar models, and
compares their simulated light curves and combined colors with the
observations. This approach greatly reduces the parameter space
over which one needs to search, and allows one to estimate the
components' masses, radii and absolute magnitudes, without
spectroscopic data. We have implemented this method in an
automated program using published theoretical isochrones and
limb-darkening coefficients. Since it is easy to automate, this
method lends itself to systematic analyses of datasets consisting
of photometric time series of large numbers of stars, such as
those produced by OGLE, MACHO, TrES, HAT, and many others surveys.
\end{abstract}

\keywords{binaries: eclipsing --- methods: data analysis --- stars: statistics --- techniques: photometric}

\section{Introduction}

Eclipsing double-lined spectroscopic binaries provide the only
method by which both the masses and radii of stars can be
estimated without having to resolve spatially the binary or rely
on astrophysical assumptions. Despite the large variety of models
and parameter-fitting implementations [e.g. WD \citep{Wilson71}
and EBOP \citep{Etzel81, Popper81}], their underlying methodology
is essentially the same. Photometric data provide the light curve
of the eclipsing binary (EB), and spectroscopic data provide the
radial velocities of its components. The depth and shape of the
light curve eclipses constrain the components' brightness and
fractional radii, while the radial velocity sets the length scale
of the system. In order to characterize fully the components of
the binary, one needs to combine all of this information. Only a
small fraction of all binaries eclipse, and spectra with
sufficient resolution and signal-to-noise can be gathered only for
bright stars. The intersection of these two groups leaves a small
number of stars.

Over the past decade, the number of stars with high-quality,
multi-epoch, photometric data has grown dramatically due to the
growing interest in finding gravitational lensing events
\citep{Wambsganss06} and eclipsing extrasolar planets
\citep{Charbonneau06}. In addition, major technical improvements
in both CCD detectors and implementations of image-difference
analysis techniques \citep{Crotts92, Alard98, Alard00} enable
simultaneous photometric measurements of tens of thousands of
stars in a single exposure. Today, there are many millions of
light curves available from a variety of surveys, such as OGLE
\citep{Udalski94}, MACHO \citep{Alcock98}, TrES \citep{Alonso04},
HAT \citep{Bakos04}, and XO \citep{McCullough06}. Despite the
increase in photometric data, there has not been a corresponding
growth in the quantity of spectroscopic data, nor is this growth
likely to occur in the near future. Thus, the number of
fully-characterized EBs has not grown at a rate commensurate with
the available photometric datasets.

In recent years, there has been a growing effort to mine the
wealth of available photometric data, by employing automated
pipelines which use simplified EB models in the absence of
spectroscopic observations and hence without a fixed physical
length scale and absolute luminosity \citep{Wyithe01, Wyithe02,
Devor04, Devor05a}. In this paper, we present a method that
utilizes theoretical isochrones and multi-epoch photometric
observations of the binary system to estimate the physical
parameters of the component stars, while still not requiring
spectroscopic observations.

Our Method for Eclipsing Component Identification\footnote{The
source code and running examples of MECI, as well as a suite of
utilities, can be downloaded from:
http://cfa-www.harvard.edu/$\sim$jdevor/MECI.html} (MECI), finds
the most probable masses, radii, and absolute magnitudes of the
stars. The input for MECI is an EB's photometric light curve and
out-of-eclipse colors (we note that in the absence of color
information, the accuracy in the estimation of the stellar
parameters is significantly reduced; \S\ref{subsecSimulate}). This
approach can be used to characterize quickly large numbers of
eclipsing binaries; however it is not sufficient to improve
stellar models, since underlying isochrones must be assumed.

In a previous paper \citep{Devor05b}, we outlined the ideas behind
both MECI and a closely related, ``quick and dirty'' alternative,
which we called MECI-express. Though MECI-express is much faster
and easier to implement, it is also far less accurate. For this
reason we will not discuss it further, and instead concentrate
exclusively on MECI. We discuss its applications
(\S\ref{secMotivation}), aspects of its implementation
(\S\ref{secMethod}), tests of its accuracy (\S\ref{secTests}),
and finally summarize our findings (\S\ref{secConclusions}).

\section{Motivation}
\label{secMotivation}

\subsection{Characterizing the binary stellar population}
\label{subsecStellarPop}

First and foremost, MECI is designed as a high throughput means to
systematically estimate the masses of large numbers of stars.
Though the result in each system is uncertain, by statistically
analyzing large catalogs, one can reduce the non-systematic
errors. Much work has already been invested into characterizing
binary systems through spectroscopic binary surveys
\citep[e.g.][]{Duquennoy91, Pourbaix04}, yet the limited data and
their large uncertainties have led to inconsistent results
\citep{Mazeh05}. The driving questions that have spurred debate in
the community include: What are the initial mass functions of the
primary and secondary components? How do they relate to the
initial mass function of single stars? What is the distribution of
the components' mass ratio, $q$, and in particular, does it peak
at unity? This lack of understanding is further highlighted by the
fact that most of the stars in our galaxy are members of binary
systems, and that these questions have lingered for over a
century. MECI may help sort this out by systematically
characterizing the component stars of many EB systems.

By requiring only photometric data, a survey using MECI can study
considerably fainter binary systems than spectroscopic surveys,
and thus remain complete to a far larger volume. As an
illustrative example, the difference image analysis of the bulge
fields of OGLE II, using the Las Campanas $1.3$m Warsaw telescope
in a drift-scan mode (an effective exposure time of $87$ seconds),
attained a median noise level of $0.1$ mag, for $I=18$ binaries,
even in moderately crowded fields \citep{Wozniak00}. In contrast
to this, the CfA digital speedometer on the $1.5$m FLWO telescope
has a spectral resolution of $R \simeq 35,000$ (at $5177$\AA) and
typically yields a radial velocity precision of $0.5 \ {\rm km \,
s^{-1}}$, with a faint magnitude limit of $V=13$ \citep{Latham92}.
Though the limiting magnitudes are very much dependent on the
throughput of the relevant instruments and the precision one
wishes to achieve, this $5$ magnitude difference for telescopes of
similar aperture corresponds to a factor of $10$ in distance or
$1000$ in volume, and illustrates the significant expansion that
can be achieved by purely photometric surveys. Conversely, one can
achieve the same magnitude limit with an aperture $10$ times
smaller. The success of this approach has been demonstrated by
several automated observatories, such as TrES \citep{Alonso04} and
HAT \citep{Bakos04}, which each use networks of observatories with
10-cm camera lenses to monitor stars to $V \simeq 13$.

\subsection{Identifying low-mass main-sequence EBs}
\label{subsecIdLowMass}

One of the most compelling applications of MECI will be to sort
quickly thousands of EBs present in large photometric surveys, and
to subsequently select a small subset of objects from the
resulting catalog for further study. In particular, lower
main-sequence stars that are partially or fully convective have
not been studied with a level of detail remotely approaching that
of solar-type (and more massive) stars. This is particularly
troubling since late-type stars are the most common in the Galaxy,
and dominate its stellar mass. It has been shown that models
underestimate the radii of low-mass stars by as much as $20$\%
\citep{Lacy77a, Torres02}, a significant discrepancy considering
that for solar-type stars the agreement with the observations is
typically within $1-2$\% \citep{Andersen98}. Similar problems
exist for the effective temperatures predicted theoretically for
low-mass stars. Progress in this area has been hampered by the
lack of suitable M-dwarf binary systems with accurately determined
stellar properties, such as mass, radius, luminosity, and surface
temperature. Detached eclipsing systems are ideal for this
purpose, but only five are known among M-type stars: CM Dra
\citep{Lacy77b, Metcalfe96}, YY Gem \citep{Kron52, Torres02}, CU
Cnc \citep{Delfosse99, Ribas03}, and OGLE BW3 V38
\citep{Maceroni97, Maceroni04}, and TrES-Her0-07621
\citep{Creevey05}. They range in mass from about $0.25 M_\sun$ (CM
Dra) to $0.6 M_\sun$ (YY Gem).  The number of such objects could
be greatly increased by using tools such as MECI to mine the
extant photometric datasets to locate these elusive low-mass
systems.

\subsection{EBs as standard candles}

Using MECI, we are able to estimate the absolute magnitude of the
binary system. Together with its extinction-corrected
out-of-eclipse apparent magnitude, we can then calculate the
distance modulus to any given EB. The estimation of distances to
EBs dates back to \citet{Stebbing10}, and their use as distance
candles in the modern astrophysical context was recently
elucidated by \citet{Paczynski97}. However, unlike these studies,
MECI does not require spectroscopy and therefore is able to
analyze binaries that are significantly less luminous (see
\S\ref{subsecStellarPop}). Though the distance estimation from
MECI will be uncertain, in many cases this will still be an
improvement over existing methods. For example, if there are many
EBs in a stellar cluster, the distance estimate can be greatly
improved by combining their results, to reduce the non-systematic
errors by a factor of the square root of the number of systems.
Following \citet{Guinan96}, one might be able to use such
clustered EB standard candles to better constrain the distance to
the LMC and SMC, and thus be able to further constrain the bottom
of the cosmological distance ladder. In the case of MECI, the
uncertainties of each distance measurement will be considerably
larger, but as suggested by Tsevi Mazeh (2005, personal
communication), this will be compensated for by the far larger
number of measurements that can be made. Another intriguing
application of such EB standard candles is to map large scale
structures in the Galaxy, such as the location and orientation of
the galactic bar, arms, and merger remnants \citep[see, for
example,][and references therein]{Vallee05}.

\section{Method}
\label{secMethod}

The EB component identification is performed in two stages. First
the orbital parameters of the EB are estimated
(\S\ref{subsecOrbitalParams}), then the most likely stellar
parameters are identified (\S\ref{subsecStellarParams}). Our
implementation of MECI has the option to fix the estimates of the
orbital parameters, or to fine-tune them for each
stellar pairing considered in the second stage. The average
running time for MECI to analyze a $1000$-point light curve on a
single $3.4$GHz Intel Xeon CPU is $0.4$ minutes. If we permit fine
tuning of the orbital parameters for each pairing, the running
time grows to $6$ minutes per light curve.

\subsection{Stage 1: Finding the orbital parameters}
\label{subsecOrbitalParams}

In the first stage, we estimate the EB's orbital parameters from
its light curve.  Many EBs have orbital periods of a few
days or less, owing to the greater probability for such systems
to present mutual eclipses, and to the limited baselines in the
datasets from which they are identified. Most of these
systems will have orbits that have been circularized due to
tidal effects. For such circular orbits, the only
parameters we seek are the orbital period, $P$, and epoch of
periastron, $t_0$. For non-circular orbits we
also fit the orbital eccentricity, $e$, and the argument of periastron,
$\omega$. The period is determined using a periodogram, and the
remaining parameters are obtained through fitting the offset,
duration and time interval between the light curve's eclipses (see
below). Holding these parameters fixed at these initial estimates
significantly reduces the computational requirements of MECI.

We postpone fitting the orbital inclination, $i$, until the second
stage, since it is difficult to determine this parameter robustly
without first assuming values for the stellar radii and masses.
This difficulty arises because it is often difficult to
distinguish a small secondary component from a large secondary
component in a grazing orbit. In stage 2, additional information,
such as the theoretical stellar mass-radius relation and colors
are used to help resolve this degeneracy.

The procedure for fitting the aforementioned parameters from the
EB light curve is a well-studied problem \citep{Kopal59, Wilson71,
Etzel91}. We chose to estimate the period with a variant of the
analysis of variances (AOV) periodogram\footnote{The source code
and running examples of both the AOV periodogram and the DEBiL
fitter can be downloaded from:
http://cfa-www.harvard.edu/$\sim$jdevor/DEBiL.html} by
\citet{SchwarzenbergCzerny89, SchwarzenbergCzerny96}. We then use
the Detached Eclipsing Binary Light curve (DEBiL)
fitter\footnotemark[3] by \citet{Devor04, Devor05a} for fitting
the remaining orbital parameters. For non-circular systems,
following \citet{Kopal59} and \citet{Kallrath99}, we estimate the
orbital eccentricity and argument of periastron from the orbital
period, the duration of the eclipses, $\Theta_{1,2}$, and the time
interval between the eclipse centers, $\Delta t$, as follows:
\begin{eqnarray}
\omega & \simeq & \arctan \left[\frac{2}{\pi} \left(\frac{\Theta_1 - \Theta_2}{\Theta_1 + \Theta_2} \right)\left(\frac{\Delta t}{P} - \frac{1}{2}\right)^{-1}\right], {\rm \ and}\\
e & \simeq & \frac{\pi}{2 \cos \omega} \left| \frac{\Delta t}{P} - \frac{1}{2} \right|.
\end{eqnarray}

In practice, it is difficult to accurately determine the eclipse
duration. We estimate this duration by first calculating
the median flux outside the eclipses, then estimating the midpoints and
depths of the eclipses using a spline. We then assign
the duration of each eclipse to be the time elapsed
from the moment at which the light curve during ingress crosses the midpoint between the
out-of-eclipse and bottom-of-eclipse fluxes, until the moment
at which the light curve crosses the corresponding point during egress.

\subsection{Stage 2: Finding the absolute stellar parameters}
\label{subsecStellarParams}

\begin{figure}
\plotone{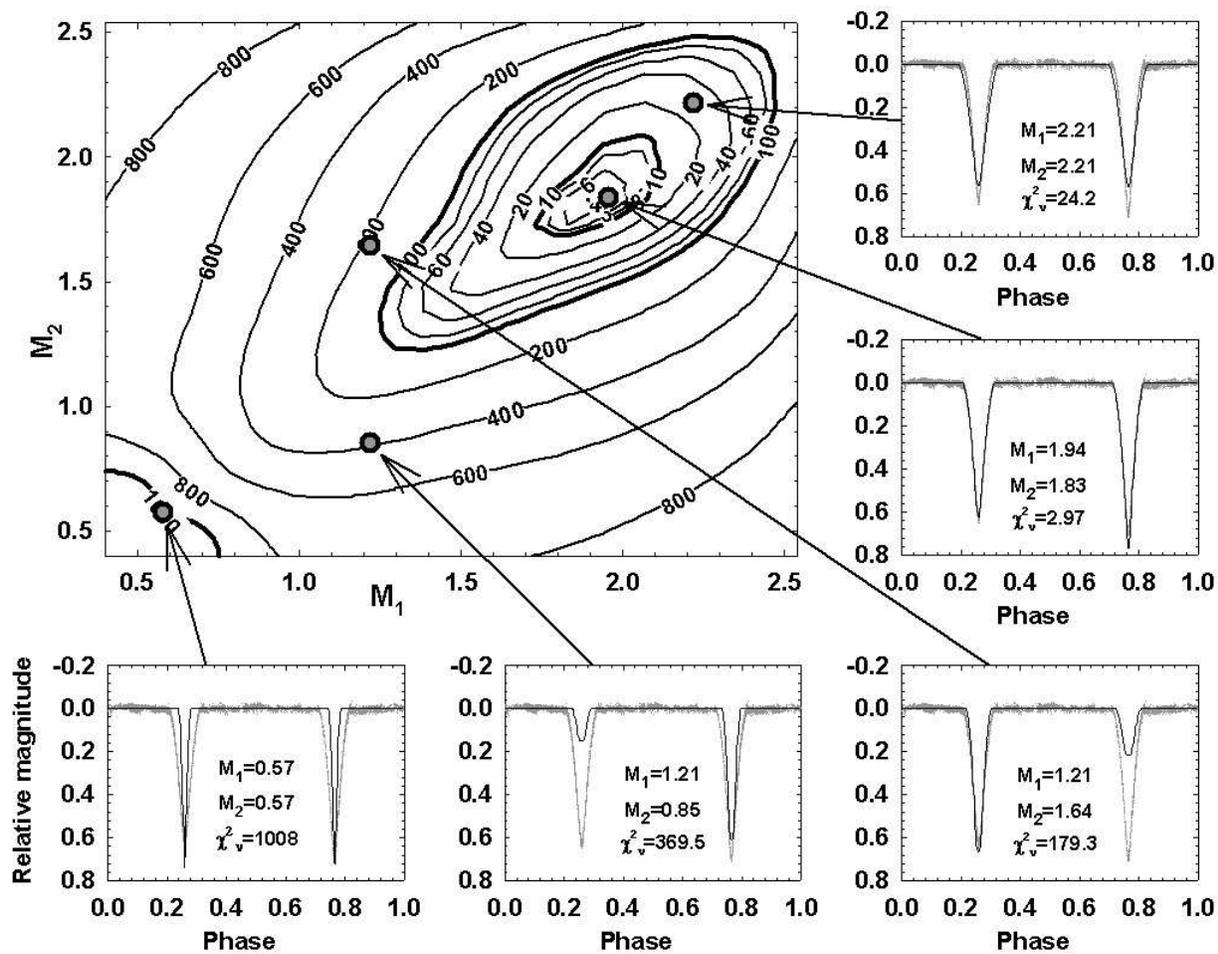}
\caption{The large upper-left panel shows the MECI $\chi^2_\nu$ surface as
a function of the assumed masses (in units of $M_{\Sun}$)
of the component stars in the WW Camelopardalis system.  The model light curve
at five locations in the grid is shown in the smaller panels, overplotted on
the observed light curve from \citep{Lacy02}.
A high-quality version of this figure can be seen at: http://cfa-www.harvard.edu/$\sim$jdevor/MECI/paper/}
\label{figMECIpanels}
\end{figure}

\begin{figure}
\plotone{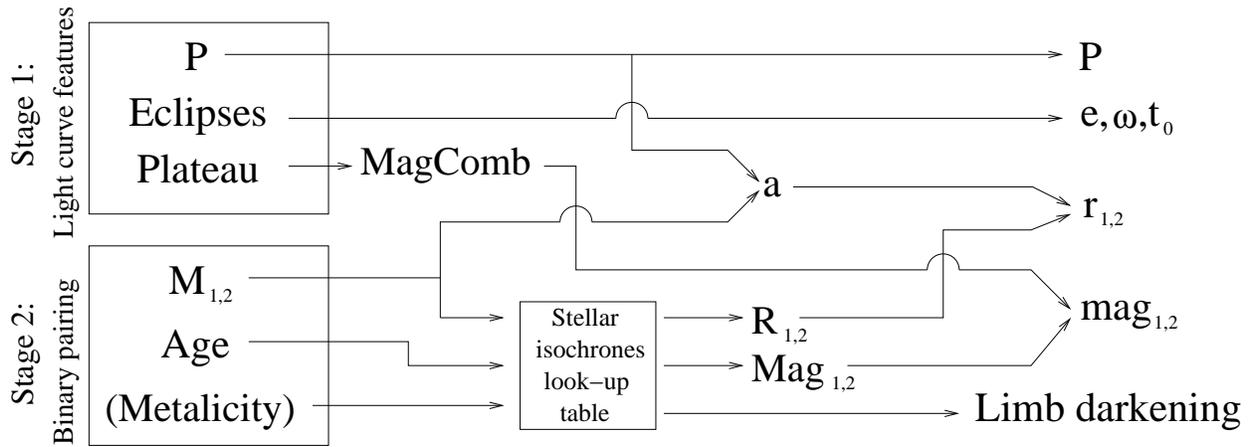}
\caption{A flow diagram demonstrating the process by which
MECI assigns the parameters to an EB based on its observed
light curve.  The details of stages~1 \& 2 are described
in \S\ref{subsecOrbitalParams} and \S\ref{subsecStellarParams},
respectively.}
\label{figMethod}
\end{figure}

In the second stage, we estimate the EB's absolute stellar
parameters by iterating through many possible stellar pairings,
simulating their expected light curves (see
Fig.~\ref{figMECIpanels}), and finding the pairing that minimizes
the $\chi^2_\nu$ function (see \S\ref{subsecLikelihoodScore}). The
parameters we fit are the masses of the two EB components,
$M_{1,2}$, their age (the components are assumed to be coeval),
and their orbital inclination, $i$. Optionally, we can also
fine-tune the orbital parameters obtained from the first stage.
This option is necessary only for binaries with eccentric orbits,
since varying their inclination will affect the fit of their
previously estimated orbital parameters.  The flow diagram for the
entire procedure is shown in Figure~\ref{figMethod}.

\placefigure{figMECIpanels}
\placefigure{figMethod}

If an estimate of the out-of-eclipse combined apparent
magnitude, $mag_{comb}$, of the EB (i.e. the light curve plateau)
is available, we may also estimate the distance modulus. If
$mag_{comb}$ is not available (for example, if the light curve has
been normalized), the distance
modulus cannot be evaluated unless an independent measurement of
the out-of-eclipse brightness is available.  In either case,
this procedure does not affect our estimates of the stellar
parameters.

Once we assume the masses and age of the binary components, we use
pre-calculated theoretical tables to
look up their absolute stellar parameters, namely their radii,
$R_{1,2}$, and absolute magnitudes, $Mag_{1,2}$.  We use the Yonsei-Yale
isochrones of solar metallicity \citep{Kim02} to specify the
binary components' radii and absolute magnitudes, in a range of
filters $(U,B,V,R,I)_{Cousins}$ and $(J,H,K)_{ESO}$.
We note that the Yonsei-Yale isochrones do not extend below
$0.4\, M_\odot$. To consider stars with masses below this value we
constructed tables from the isochrones of \citet{Baraffe98}, which
are generally more reliable for masses below $0.75\, M_\odot$.

Together with the orbital parameters
(\S\ref{subsecOrbitalParams}), we have all the information
required to simulate the EB light curve. The fractional radii,
$r_{1,2}$, and apparent magnitudes, $mag_{1,2}$, of the binary
components, which are needed for this calculation, are calculated
as follows:
\begin{eqnarray}
a &=& [G(M_1 + M_2)(P/2\pi)^2]^{1/3} \simeq\\
  & & 4.206 R_{\odot}(M_1/M_{\odot} + M_2/M_{\odot})^{1/3} (P/day)^{2/3}\nonumber,\\
r_{1,2} &=& R_{1,2} / a,\\
mag_1 &=& mag_{comb} + 2.5\log \left[1 + 10^{-0.4(Mag_2 - Mag_1)}\right], {\rm \ and}\\
mag_2 &=& mag_1 + (Mag_2 - Mag_1).
\end{eqnarray}

We create model light curves using DEBiL, which has a fast light
curve generator. DEBiL assumes that the EB is detached, with
limb-darkened spherical components (i.e. no tidal distortions or
reflections). To describe the stellar limb darkening, it employs
the quadratic law \citep{Claret95}:
\begin{equation}
I(\theta ) = I_0 \left[ {1-\tilde{a}(1-\cos \theta)-\tilde{b}(1-\cos \theta )^2} \right],
\end{equation}
where $\theta$ is the angle between the line of sight and the
emergent flux, $I_0$ is the flux at the center of the stellar
disk, and $\tilde{a}$, $\tilde{b}$ are coefficients that define
the amplitude of the center-to-limb variations.  We use the ATLAS
\citep{Kurucz92} and PHOENIX \citep{Claret98, Claret00} tables to
look up the quadratic limb-darkening coefficients, for high-mass
($T_{eff} \geq 10000$K or $\log g \leq 3.5$) and low-mass
($T_{eff} < 10000$K and $\log g > 3.5$) main-sequence stars
respectively.

Finally, the orbital inclination is fit at each iteration so to
make the simulated light curve most similar to the observations.
For this we employed the robust ``golden section'' bracket search
algorithm \citep{Press92}. This inner loop dominates the
computational time required. In the case of non-circular orbits,
it is often necessary to iterate the estimates of the orbital
parameters ($e$, $t_0$, $\omega$, $i$). When this option is
enabled, MECI employs the rolling simplex algorithm
\citep{Nelder65, Press92}, which fits all four orbital parameters
simultaneously.

\placefigure{figMECIpanels}

\subsection{Assessing the likelihood of a binary pairing}
\label{subsecLikelihoodScore}

The observational data for each EB consists of $N_{lc}$ observed
magnitudes $O_i$, each with an associated uncertainty $\epsilon_i$,
as well as $N_{colors}$ out-of-eclipse colors $\tilde{O}_c$,
each with an uncertainty $\tilde{\epsilon}_c$.  Our model
yields the corresponding predicted light curve magnitudes
$C_i$ and out-of-eclipse colors $\tilde{C}_c$.
We define the goodness-of-fit function to be:
\begin{equation}
\chi^2_\nu = \frac{1}{w+N_{colors}}\left[\frac{w}{N_{lc}}\sum_{i=1}^{N_{lc}} \left(\frac{O_i - C_i}{\epsilon_i}\right)^2 +
\sum_{c=1}^{N_{colors}} \left(\frac{\tilde{O}_c - \tilde{C}_c}{\tilde{\epsilon}_c}\right)^2\right],
\end{equation}
where $w$ is a factor that describes the relative weights assigned
to the light curve and color data (see below). The value of
$\chi^2_\nu$ should achieve unity if the assumed model accurately
describes the data, and that the errors are Gaussian-distributed
and are estimated correctly.

In practice, typical light curves may have $N_{lc} > 1000$ points,
whereas only $1 \leq N_{colors} \leq 5$ might be available. We
have found it necessary to select a value for $w$ that increases
the relative weight of the color information to obtain reliable
results ($w < N_{lc}$). In general, the optimal value for $w$ will
depend on the accuracy of the observed colors $\tilde{O}_c$ and
the degree to which the EB light curve deviates from the
assumption of two well-detached, limb-darkened spherical
components. Based on the tests described in \S\ref{secTests}, we
find that a wide range of values for $w$ produces similar results,
and that values in the range $10 \leq w \leq 100$ most accurately
recover the correct values for the stellar parameters.

We identify the global minimum of $\chi^2_\nu$ in three steps:
First, we calculate the value of $\chi^2_\nu$ at all points in a
coarse $N \times N$ grid at each age slice. The $N$ mass values
are selected to be spaced from the lowest mass value present in
the models to the greatest values at which the star has not yet
evolved off the main sequence. Next, we identify any local minima,
and refine their values by evaluating all available intermediate
mass pairings. Finally, we identify the global minimum from the
previous step, and fit an elliptic paraboloid to the local
$\chi^2_\nu$ surface around the lowest minimum.  We assign the
most likely values for the stellar masses and age to be the
location of the minimum of the paraboloid. The curvature of the
paraboloid in each axis provides the estimates of the
uncertainties in these parameters. In practice, these formal
uncertainties underestimate the true uncertainties since they do
not consider the systematic errors due to (1) the over-simplified
EB model, (2) errors in the theoretical stellar isochrones and
limb-darkening coefficients, and (3) sources of non-Gaussian
noise in the data.

When choosing the value of $N$ above, we must balance
computational speed considerations with the risk of missing the
global minimum by under-sampling the $\chi^2_\nu$ surface. For
most main-sequence EBs, the $\chi^2_\nu$ surface contains only
one, or at most a few local minima, and our experience is that
$N=10$ usually suffices (see \S\ref{subsecSimulate}). For systems
that are either very young or in which a component has begun to
evolve off the main-sequence, the $\chi^2_\nu$ surface requires a
much denser sampling. Evolved components, which may be present in
as many as a third of the EBs of a magnitude-limited phorometric
survey \citep{Alcock97}, introduce an additional challenge if
their isochrones intersect other isochrones on the color-magnitude
diagram. At such intersection points, stars of different masses
will have approximately equal sizes and effective temperatures,
creating degenerate regions on the $\chi^2_\nu$ surface. This
degeneracy can, in principal, be broken with sufficient color
information, which will probe differences in the stars' limb
darkening and absorption features, both of which vary with surface
gravity.

We also note that multiple local minimum may result for light curves
with very small formal uncertainties.  In this case, numerical
errors in the simulated light curve dominate.  This problem can
be mitigated by increasing the number of
iterations used in fitting the orbital parameters (see
\S\ref{subsecStellarParams}).

\subsection {Optimization}
\label{secOptimization}

We implemented a number of optimizations to increase the speed of
MECI. First, since each light curve is independent, we parsed the
data set and ran MECI in parallel on multiple CPUs. Second, we
reduced the number of operations by identifying and skipping
unphysical stellar pairings. Specifically, we required ($r_1 + r_2
< 0.8$) to preclude binaries that were not well detached.  In
addition, for EBs with clear primary and secondary eclipses, we
skipped high-contrast-ratio pairings for which the maximum depth
of the primary eclipse, $\Delta mag_{1}$, or the maximum depth of
the secondary eclipse, $\Delta mag_{2}$, fell below a specified
threshold, $\Delta mag_{\rm cutoff}$. In particular, we skipped
over pairings for which $\min \left( {\Delta mag_{1}}, {\Delta
mag_{2}} \right) \le {\Delta mag_{\rm cutoff}}$, where
\begin{eqnarray}
\Delta mag_{1} &\simeq& 2.5 \log \left[1 + \frac{(R_2/R_1)^2}{1 - (R_2/R_1)^2 + 10^{0.4(Mag1 - Mag2)}}\right], {\rm \ and}\\
\Delta mag_{2} &=& 2.5 \log \left[1 + 10^{0.4(Mag1 - Mag2)}\right].
\end{eqnarray}

These estimates assume equatorial eclipses, since we
seek to evaluate the maximum possible eclipse depths.
The first expression is approximate because it neglects
the effects of limb-darkening on the eclipse depth.
In practice, the chosen value for $\Delta mag_{\rm cutoff}$
will depend on the typical precision and cadence of the
data set in question.

We note here a special case that we shall revisit in \S\ref{secPitfalls}.
For EB light curves with equally spaced eclipses of equal depth,
we must also consider the possibility that our assumed period is double
the true value, and hence the secondary eclipse is undetected.
When we identified such cases, we analyzed the light curve
as usual but removed the above requirement.  In such cases,
we can place only an upper limit on the mass of the secondary component.

\section {Testing MECI}
\label{secTests}

In order to establish the accuracy and reliability of MECI under a
variety of scenarios, we conducted two distinct tests.

\subsection {Observed Systems}
\label{ObservedSystems}

The first test was to run MECI on several observed
light curves of eclipsing binary systems whose stellar parameters
had been precisely determined from detailed
photometric and spectroscopic studies.

\begin{figure}
\plotone{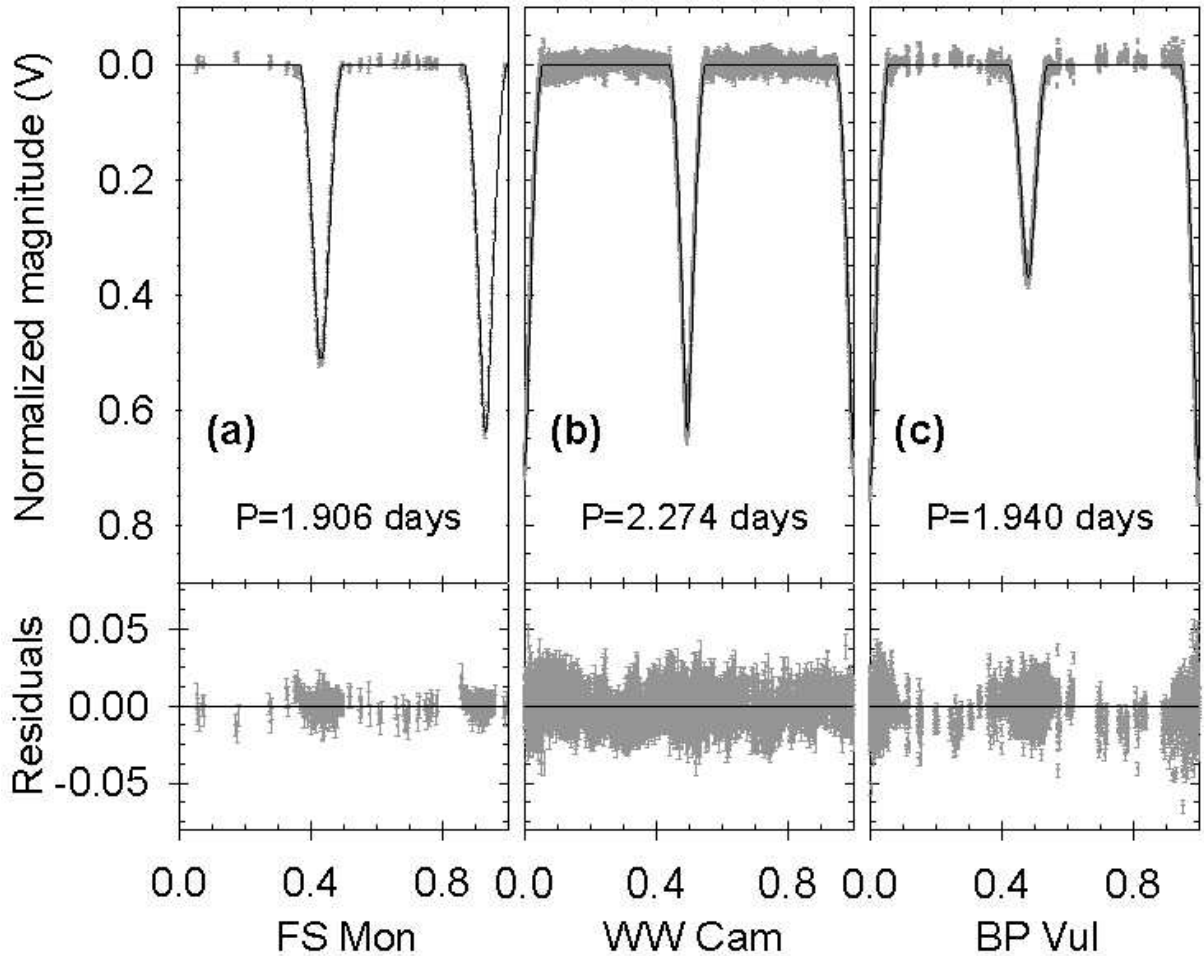}
\caption{The observed light curves of FS~Monocerotis (Lacy et al.\ 2000), WW~Camelopardalis
(Lacy et al.\ 2002), and BP~Vulpeculae (Lacy et al.\ 2003), each overplotted with the
best-fit model DEBiL solution used in our MECI algorithm.  The masses
and ages corresponding to these solutions are listed in Table~\ref{tableLacy}. The residuals
to each fit are shown in the lower panels. A high-quality version of this figure can be seen at:
\newline http://cfa-www.harvard.edu/$\sim$jdevor/MECI/paper/}
\label{figLacyLC}
\end{figure}

\begin{figure}
\plotone{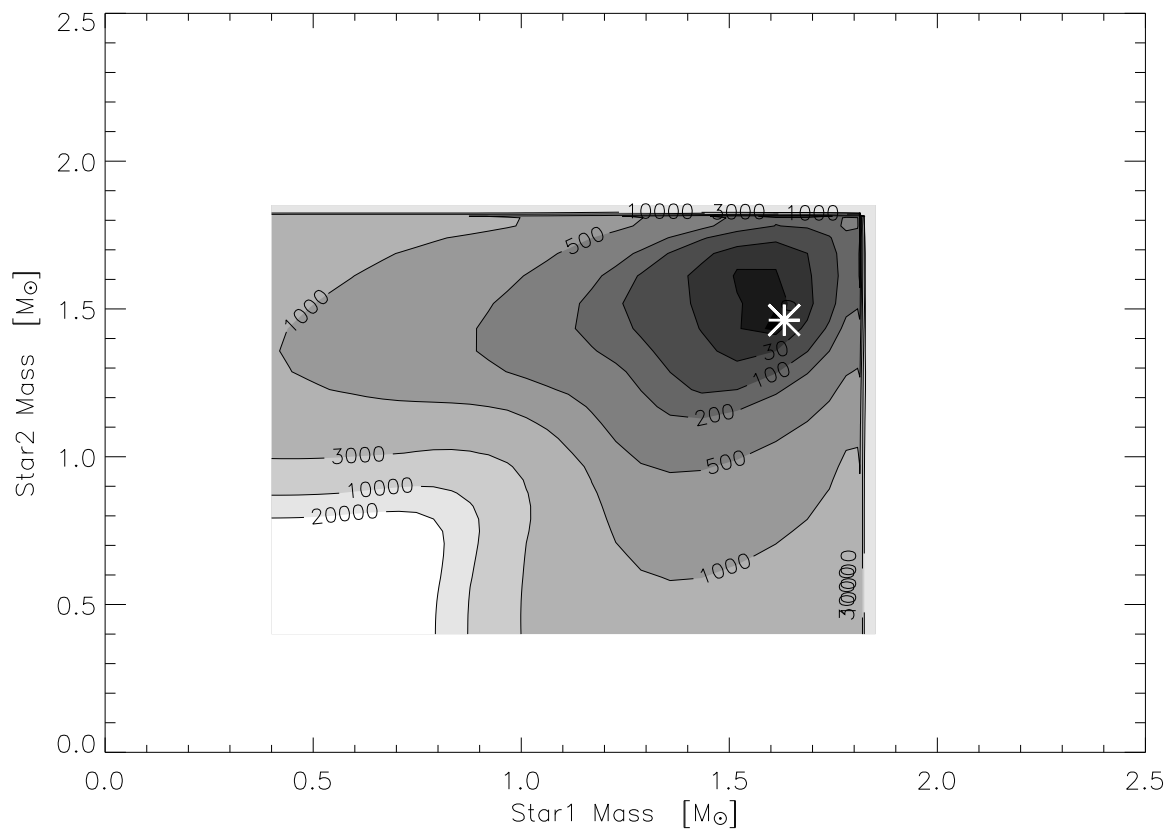}
\caption{The MECI $\chi^2_\nu$ surface to the FS~Monocerotis light curve and colors (Lacy et al.\ 2000),
assuming an age of $1.6$~Gyr and fixing $w=10$. The estimate of the stellar masses (Lacy et al.\ 2000)
from a combined analysis of the light curve and spectroscopic observations is indicated by
a white asterisk, and is near to the minimum identified by MECI.
Note the erratic behavior of the contours at the upper end of the mass range, which
results from the rapid evolution of stars of those masses at this age.}
\label{figLacy00}
\end{figure}

\begin{figure}
\plotone{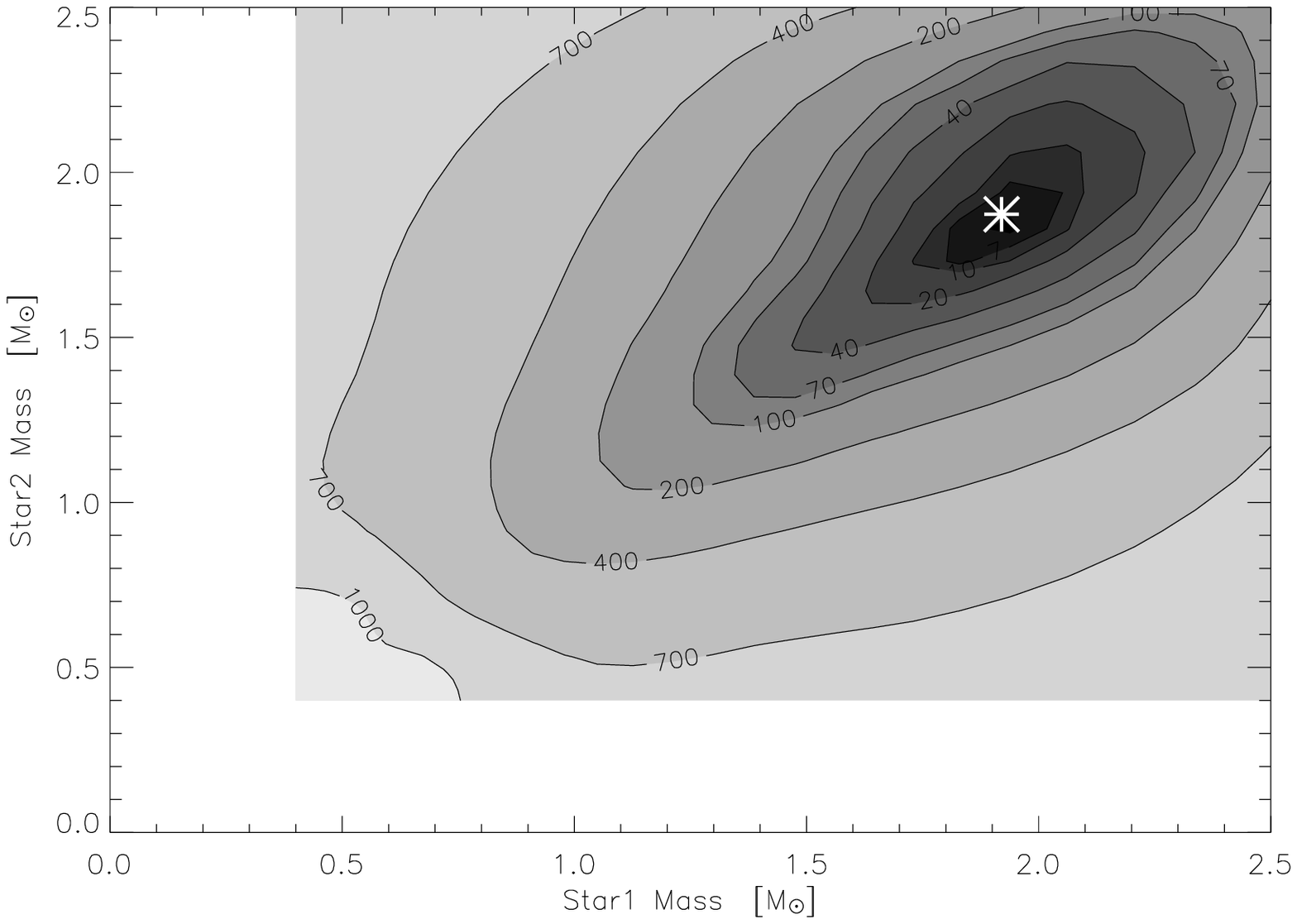}
\caption{The MECI $\chi^2_\nu$ surface to the WW~Camelopardalis light curve and colors (Lacy et al.\ 2002),
assuming an age of $0.6$~Gyr and fixing $w=10$. The estimate of the stellar masses (Lacy et al.\ 2002)
from a combined analysis of the light curve and spectroscopic observations is indicated by
a white asterisk, and is extremely close to the solution identified by MECI.}
\label{figLacy02}
\end{figure}

\begin{figure}
\plotone{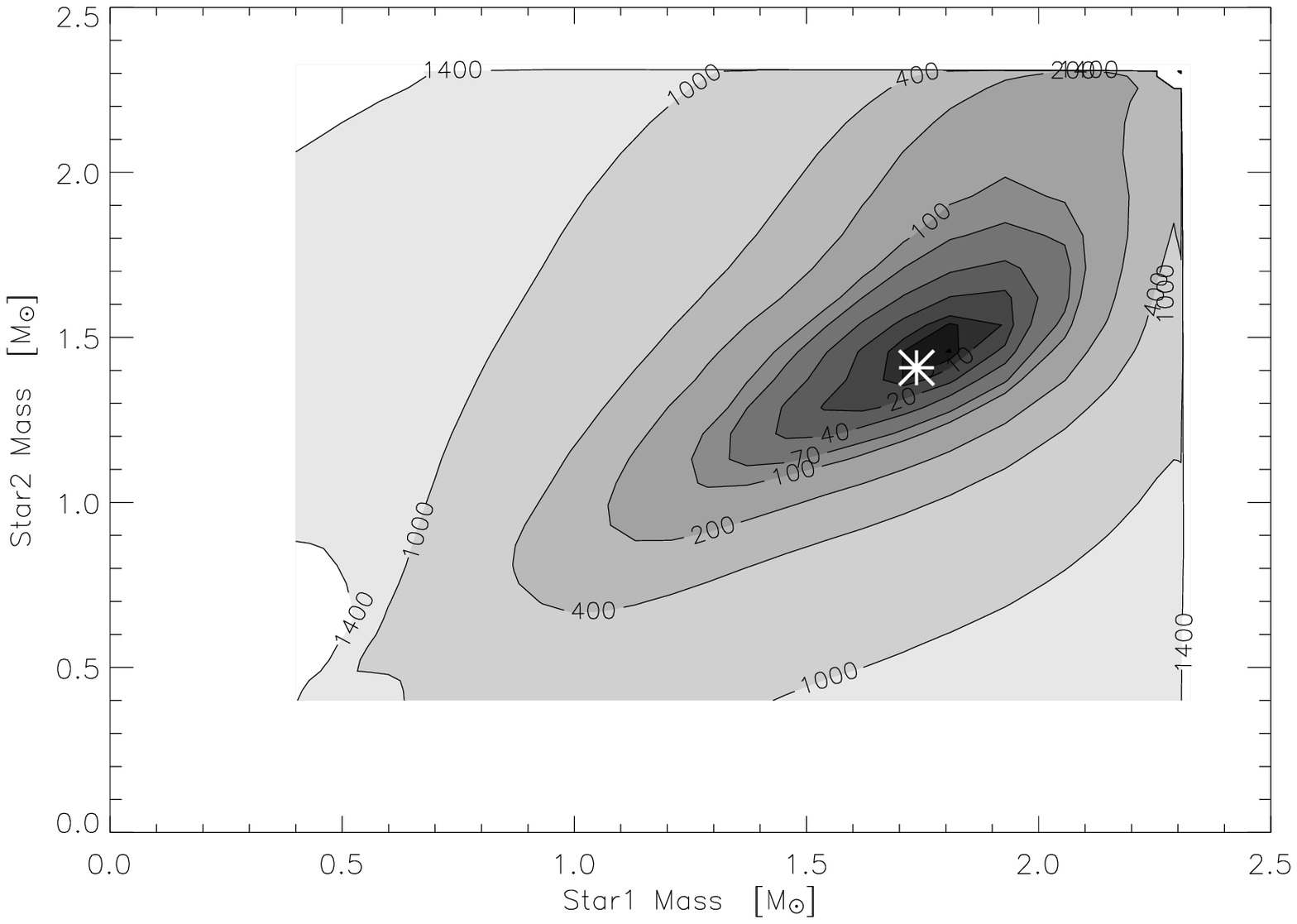}
\caption{The MECI $\chi^2_\nu$ surface to the BP~Vulpeculae light curve and colors (Lacy et al.\ 2003),
assuming an age of $0.8$~Gyr and fixing $w=10$. The estimate of the stellar masses (Lacy et al.\ 2003)
from a combined analysis of the light curve and spectroscopic observations is indicated by
a white asterisk, and is extremely close to the solution identified by MECI.}
\label{figLacy03}
\end{figure}

We examined three well-studied EBs. The first was FS Monocerotis
\citep{Lacy00}, for which we modeled the published light curve,
which had $N_{lc}=249$ data points, as well as the published $U-B$
and $B-V$ colors. The second was WW Camelopardalis \citep{Lacy02},
for which we modeled the published light curve, which had
$N_{lc}=5759$ observations, as well as the $B-V$ color.  Finally,
we studied BP Vulpeculae \citep{Lacy03}, for which we modeled the
published light curve, which had $N_{lc}=5236$ observations, as
well as the $B-V$ color. All three published light curves were
observed in $V$-band and are plotted in Figure~\ref{figLacyLC}.
The colors had been corrected for reddening. The contour plots of
the ${\chi}^2_\nu$ surfaces resulting from our MECI analysis
(setting the weighting $w=10$) are shown in
Figures~\ref{figLacy00}, \ref{figLacy02}, and \ref{figLacy03}.
Note that FS~Mon is more tightly constrained due to its greater
color information. Furthermore, the asymmetry in BP Vol's contour
is due to its unequal eclipse depths. In all cases, the
${\chi}^2_\nu$ surface has a single minimum, which is close to the
published values. In Table~\ref{tableLacy}, we tabulate the
results of our analysis and compare these to the published values.

We then changed the weighting factor to $w=100$ and repeated this
procedure. The MECI results for FS~Mon and BP~Vul were essentially
identical to our earlier findings for $w=10$. In the case of
WW~Cam, the results for $w=10$ were significantly closer to the
published values. This is likely due to the fact that it is a
young system ($age=500$~Myr), for which the brightness and radii
at constant mass vary significantly. Thus, the lower light curve
information weighting brought about smoother $\chi^2_\nu$ contours
(see \S\ref{subsecLikelihoodScore}).

\placefigure{figLacyLC}
\placefigure{figLacy00}
\placefigure{figLacy02}
\placefigure{figLacy03}

\subsection {Simulated systems}
\label{subsecSimulate}

In our second test, we produced large numbers of simulated EB
light curves with various levels of injected noise, and
subsequently analyzed these photometric datasets with MECI. We
then compared the input and derived estimates of the stellar
masses and ages in order to quantify the accuracy of the MECI
analysis.

We selected the orbital and stellar parameters of each simulated
EB as follows. First, we drew an age at random from a uniform
probability distribution between $200$~Myr and $10$~Gyr. We then
selected the masses of the two EB components independently from a
flat distribution from $0.4M_{\odot}$ and the maximum mass at
which stars of this age would still be located on the
main-sequence. We then assigned the orbital period by drawing a
number from a uniform probability distribution spanning $0 < P
\le 10$~days. Similarly, we assigned the epoch of perihelion by
drawing from a uniform probability distribution spanning $0 \le
t_0 < P$, and the orbital inclination from a uniform
distribution within the range that produces eclipses,
$\arccos(r_1+r_2) \le i \le {\pi}/2$. For the tests of eccentric
systems, we also randomly selected an eccentricity, uniformly from
$0 \le e \le 0.1$, and randomly selected the angle of perihelion,
uniformly from $0 \le \omega < 2{\pi}$. Finally, we rejected any
EB system if its components were overlapping or in near contact,
$r_1 + r_2 \geq 0.8$. We also filtered out EBs with undersampled
eclipses, or for which one of the eclipse depths was smaller than
the assumed $1 \, \sigma$ noise level.

Each simulated light curve contained 1000 $R$-band data points, to
which we injected Gaussian-distributed noise. When color
information was required, we computed the out-of-eclipse
photometric colors for each EB, and injected a $0.02$~mag
Gaussian-distributed error to this value. The colors we considered
were $(V-I)_{Cousins}$, which is similar to the color provided by
the OGLE II catalog \citep{Wozniak02}, as well as $(J-H)_{ESO}$
and $(H-K)_{ESO}$, which are similar to the colors provided by the
2MASS catalog\footnote{The 2MASS catalog uses custom J, H, and
$K_s$ filters, which can be approximately converted to the ESO
standard using linear transformations \citep{Carpenter01}.}
\citep{Kleinmann94}.

We simulated 8 sets of 2500 systems each,
with the sets differing in the following respects
(see Table~\ref{tableSimulations}):  (1) circular or
eccentric orbits, (2) the number of points in the search
grid, (3) the value of $w$, which describes the relative
weight between the color and photometric data, and (4) the
availability of color information.

\begin{figure}
\epsscale{.8}\plotone{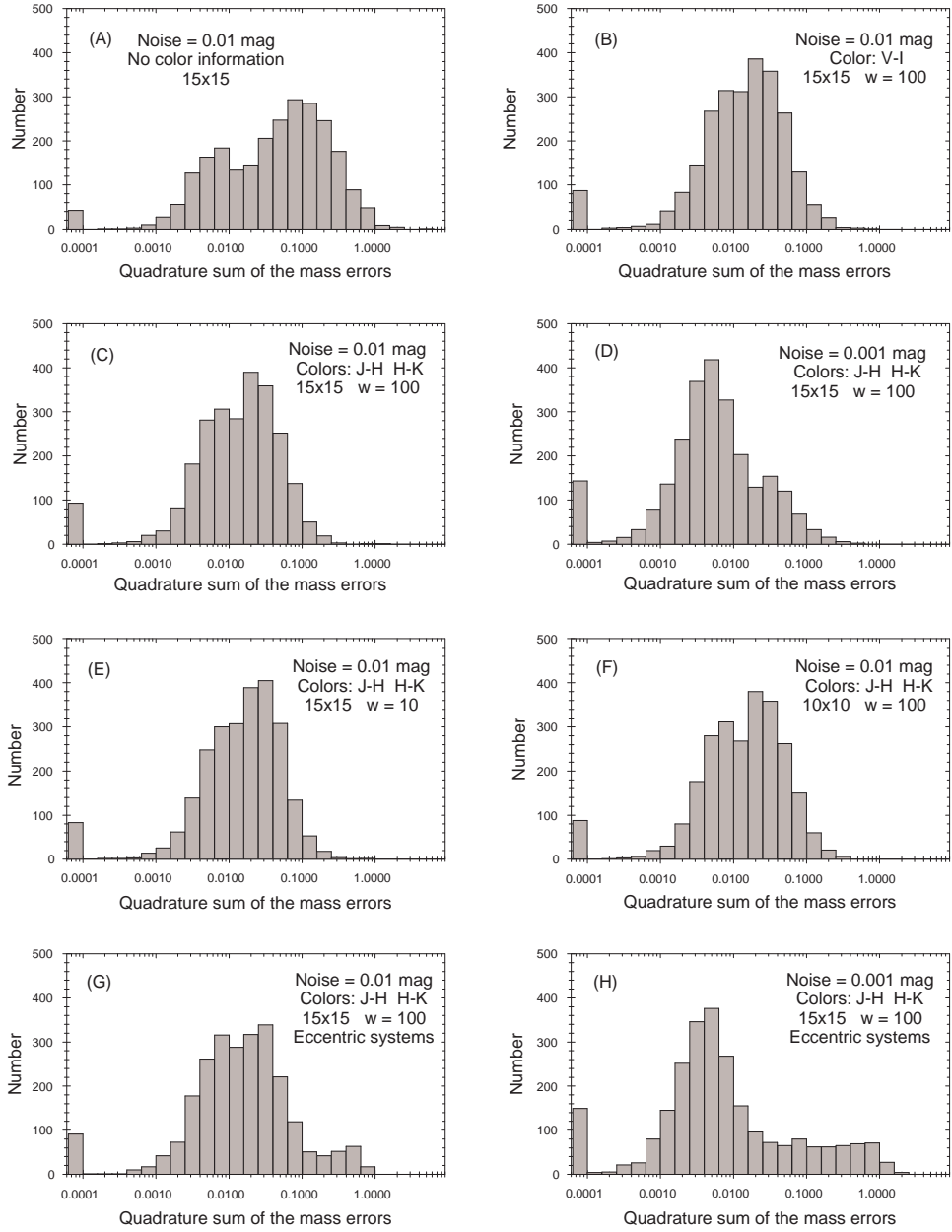}
\caption{Each panel shows the histogram of the
quadrature sum of the relative differences in the assumed and
calculated masses for the stellar components, for each of the
sets of simulated light curves described in
Table~\ref{tableSimulations}. Each set contains $2500$ simulated
EBs as described in \S\ref{subsecSimulate}, and the key parameters
of each set are listed in the upper right corner of each panel.
The leftmost bin contains the sum of all results with values less
than 0.0001. The ability of the method to accurately assign the
masses to the component stars degrades significantly in the
absence of any color information (upper left panel), but is
generally robust against changes in the particular choice of $w$
or $N$ (see \S\ref{subsecSimulate}).}
\label{figHist}
\end{figure}

In order to summarize the accuracy of the MECI results,
we computed the quadrature sum of the relative differences between
the assumed and derived values for the masses of the two components.
We plot the histograms of these values in Figure~\ref{figHist}.
In each histogram, we identify the value encompassing the region
that contains 90\% of the results.  We call
this range the ``$90^{th}$ percentile error'', and list
it in the final column of Table~\ref{tableSimulations}.

We find that the inclusion of color information significantly
improves the accuracy of the MECI results, lowering the $90^{th}$
percentile error from $30$\% in set (A), to less than $6$\% in
sets (B) and (C). In contrast, changing the value of $w$ from
$100$ in set (C), to $10$ in set (E), results in only a modest
increase of $0.8$\% in the size of the $90^{th}$ percentile error.
This indicates that the results are robust to the particular
choice of $w$.  We note, however, that a value of $w>100$ will
usually provide too little weight to the color information, which
results in poorer accuracy. An extreme example of this is seen in
set (A).

Similarly, MECI is not sensitive to the exact value of the search
grid size.   In particular, decreasing the grid size from $15
\times 15$ in set (C), to $10 \times 10$ in set (F), increases the
$90^{th}$ percentile error only modestly, from $5.8$\% to $6.1$\%.
This stability results from the fact that the $\chi^2_\nu$
function contains a broad minimum, which is well sampled even with
$N=10$ grid points.  We note, however, that this is no longer the
case when considering evolved star systems (e.g.
\S\ref{subsecLikelihoodScore}), for which a larger number of grid
points is required.

When we decreased the level of the noise injected into
the photometric time series from $0.01$ mag in set (C),
to $0.001$ mag in set (D), the $90^{th}$ percentile error dropped
from $5.8$\% to $4.0$\%.
Surprisingly, the tail of the upper end of the error distribution
extends to larger values in set (D). This appears to be due to the
phenomenon discussed in \S\ref{subsecLikelihoodScore}, whereby the
$\chi^2_\nu$ function occasionally contains many local minima.
This problem becomes acute for eccentric
systems, since they have a far more complex $\chi^2_\nu$ function.
Decreasing their noise from $0.01$ mag in set (G), to $0.001$ mag
in set (H), raises the $90^{th}$ percentile error from
$8.8$\% to $23$\%. This relatively poor performance reflects the
algorithm's inability to robustly identify the global minimum
under these conditions. In such cases one must increase the size
of the search grid and iteratively solve for the orbital parameters
of the systems, which results in a
significant increase in the computational time.

\placefigure{figHist}

\subsection{Limitations}
\label{secPitfalls}

A significant degeneracy results for light curves in which two
distinct eclipses are not apparent.  For such systems, two
distinct possibilities exist, namely that either the EB consists
of two twin components with an orbital period $P$, or that the EB
consists of two stars with very disparate sizes (such that the
secondary eclipse is not discernable), with an orbital period $2\,
P$.  It is often necessary to flag such systems and conduct
analyses with both possible values for the orbital periods.
Distinguishing which of these possibilities is the correct
solution is challenging, but in some instances there are clues.
One such clue is a variable light curve plateau that results from
the mutual tidal distortions, which in turn might indicate the
true orbital period (twice that of the observed modulation).  A
second possibility is a red excess in the system color indicating
a low-mass secondary.  Of course, follow-up spectroscopic
observations can readily resolve this degeneracy, either by
indicating the presence of two components of similar brightness,
or through a direct determination of the orbital period.

We note that MECI employs a simplified model for the generation of
the light curves (DEBiL), which can bring about additional
complications when applied to systems in which our assumptions
(see \S\ref{subsecStellarParams}) do not hold. For example, our
model ignores the effect of third light, either from a physically
associated star or a chance superposition, which reduces the
apparent depths of the eclipses and may contaminate the estimate
of the system color. Furthermore, we have ignored reflection
effects, which can raise the light curve plateau at times
immediately preceding or following eclipses. Finally, tidal
distortions will increase the apparent system brightness at
orbital quadrature, which can serve to increase the apparent depth
of the eclipses. In order for MECI to be able to properly handle
these cases, its light curve generator must be replace with a more
sophisticated one (e.g. WD or EBOP), which will likely make MECI
significantly more computationally expensive.

\section{Conclusions}
\label{secConclusions}

We have described a method for identifying an EB's components
using only its photometric light curve and combined colors. By
utilizing theoretical isochrones and limb-darkening coefficients,
this method greatly reduces the EB parameter space over which one
needs to search. Using this approach, we can quickly estimate the
masses, radii and absolute magnitudes of the components, without
spectroscopic data. We described an implementation of this method,
which enables the systematic analyses of datasets consisting of
photometric time series of large numbers of stars, such as those
produced by OGLE, MACHO, TrES, HAT, and many others.  Such
techniques are expected to grow in importance with the next
generation surveys, such as Pan-STARRS \citep{Kaiser02} and LSST
\citep{Tyson02}.  In a future publication, we shall describe a
specific application of these codes, namely to search for low-mass
eclipsing binaries in the TrES dataset.

\acknowledgments

We would like to thank Guillermo Torres for many useful
discussions and critiques, and we would like to thank Tsevi Mazeh
for sharing his ideas regarding applications for this method. We
would also like to thank Sarah Dykstra for her help editing this
paper.

{}

\clearpage

\begin{deluxetable}{lccc|ccc|ccc}
\tabletypesize{\scriptsize}
\tablecaption{Accuracy of MECI parameter estimates for 3 well-studied binaries.}
\tablewidth{0pt}
\tablehead{ & \multicolumn{3}{c}{MECI ($w=10$)} & \multicolumn{3}{c}{MECI ($w=100$)} & \multicolumn{3}{c}{\citet{Lacy00, Lacy02, Lacy03}}}
\startdata
System  & Mass 1        & Mass 2        & Age     & Mass 1        & Mass 2        & Age     & Mass 1        & Mass 2        & Age     \\
        & $[M_{\odot}]$ & $[M_{\odot}]$ & $[Gyr]$ & $[M_{\odot}]$ & $[M_{\odot}]$ & $[Gyr]$ & $[M_{\odot}]$ & $[M_{\odot}]$ & $[Gyr]$ \\
\hline
FS Monocerotis    & 1.58     & 1.47     &  1.6    & 1.57     & 1.47     &  1.6    & 1.632      & 1.462      & 1.6      \\
($N_{lc}=249$)   & [3.3\%]  & [0.5\%]  & [0.3\%] & [3.6\%]  & [0.5\%]  & [0.1\%] & $\pm0.012$ & $\pm0.010$ & $\pm0.3$ \\
WW Camelopardalis & 1.92     & 1.86     &  0.5    & 2.10     & 2.02     &  0.4    & 1.920      & 1.873      & 0.5      \\
($N_{lc}=5759$)  & [0.2\%]  & [0.9\%]  & [3\%]   & [9.6\%]  & [8.0\%]  & [17\%]  & $\pm0.013$ & $\pm0.018$ & $\pm0.1$ \\
BP Vulpeculae     & 1.78     & 1.48     &  0.7    & 1.77     & 1.48     &  0.8    & 1.737      & 1.408      & 1.0      \\
($N_{lc}=5236$)  & [2.2\%]  & [5.3\%]  & [26\%]  & [1.9\%]  &
[5.2\%]  & [22\%]  & $\pm0.015$ & $\pm0.009$ & $\pm0.2$
\enddata
\tablecomments{The rightmost columns list the masses, ages, and errors
of the component stars as determined by a combined analysis
of their light curves and spectroscopic orbits \citep{Lacy00, Lacy02, Lacy03}.
The leftmost columns list the estimates of these
quantities produced by MECI assuming $w=10$, and the central
columns list the estimates from MECI assuming $w=100$.
The square brackets indicate the fractional errors of the MECI results
with respect to the numbers in the rightmost columns.}
\label{tableLacy}
\end{deluxetable}

\begin{deluxetable}{llccclc}
\tabletypesize{\scriptsize}
\tablecaption{Accuracy of MECI mass estimates for simulated systems.}
\tablewidth{0pt}
\tablehead{Set & Noise & Orbit & Search grid & Weighting & Color information & $90^{th}$ percentile error}
\startdata
A & 0.01  mag & circular  & $15 \times 15$ & N/A    & No color information            & 30\%  \\
B & 0.01  mag & circular  & $15 \times 15$ & $w=100$  & $(V-I)_{Cousins}$               & 5.9\% \\
C & 0.01  mag & circular  & $15 \times 15$ & $w=100$  & $(J-H)_{ESO}$ and $(H-K)_{ESO}$ & 5.8\% \\
D & 0.001 mag & circular  & $15 \times 15$ & $w=100$  & $(J-H)_{ESO}$ and $(H-K)_{ESO}$ & 4.0\% \\
E & 0.01  mag & circular  & $15 \times 15$ & $w=\ 10$ & $(J-H)_{ESO}$ and $(H-K)_{ESO}$ & 6.6\% \\
F & 0.01  mag & circular  & $10 \times 10$ & $w=100$  & $(J-H)_{ESO}$ and $(H-K)_{ESO}$ & 6.1\% \\
G & 0.01  mag & eccentric & $15 \times 15$ & $w=100$  & $(J-H)_{ESO}$ and $(H-K)_{ESO}$ & 8.8\% \\
H & 0.001 mag & eccentric & $15 \times 15$ & $w=100$  & $(J-H)_{ESO}$ and $(H-K)_{ESO}$ & 23\%
\enddata
\tablecomments{The parameters of the 8 distinct sets of
simulated EB light curves that we generated and subsequently
analyzed with MECI.  The rightmost column lists the range of the
quadrature sum of the fractional errors on the masses which
encompasses 90\% of the solutions (see Fig.~\ref{figHist}), which
we take to be indicative of the accuracy of MECI under the
specified conditions.}
\label{tableSimulations}
\end{deluxetable}

\clearpage

\end{document}